\documentclass[twocolumn, table]{aastex63}
\usepackage{xcolor}
\usepackage[utf8]{inputenc}
\usepackage{comment}
\usepackage{gensymb}

\usepackage{amsmath}
\usepackage{amssymb}
\usepackage{braket}
\usepackage{amssymb}
\usepackage{color}
\usepackage{xcolor}
\usepackage{graphicx}
\usepackage{latexsym}
\usepackage{listings}
\usepackage{mathrsfs}
\usepackage{letltxmacro}
\usepackage[normalem]{ulem}

\usepackage{subfigure}

\usepackage{verbatim}
\usepackage{tabularx}
\usepackage{ragged2e}
\usepackage{multirow}
\usepackage{amsbsy}

\usepackage{numprint}
\usepackage{makecell}
\usepackage[utf8]{inputenc}

\newcommand{\R}{\mathbb{R}}
\newcommand{\N}{\mathbb{N}} 

\newcommand{\V}{\mathcal{V}}

\newcommand{\F}{\mathcal{F}}

\newcommand{\M}{\mathcal{M}}

\newcommand{\I}{\mathcal{I}}

\newcommand{\B}{\text{B}}

\npdecimalsign{.}
\npthousandsep{,}

\newcolumntype{Y}{>{\RaggedRight\arraybackslash}X}

\graphicspath{{./figures/}}

\sloppypar

\received{June 1, 2019}
\revised{January 10, 2019}
\accepted{\today}
\submitjournal{RNAAS}

\shorttitle{COSMIC v MESA Comparison}
\shortauthors{Kenoly, Luu, et al.}
\graphicspath{{./}{figures/}}

\begin{document}

\title{Understanding Binary Systems - a Comparison between COSMIC and MESA}

\author{Lailani Kenoly}
\affiliation{De Anza College, Cupertino, CA 95014}
\affiliation{CCS-2, Computational Physics and Methods, Los Alamos National Lab, Los Alamos, NM 87544}
\author{A. Ken Luu}
\affiliation{Department of Physics, San Francisco State University, San Francisco, CA 95132}
\affiliation{CCS-2, Computational Physics and Methods, Los Alamos National Lab, Los Alamos NM 87544}
\author{Celia Toral}
\affiliation{Department of Physics, Cornell University, Ithaca, NY 14850}
\affiliation{T-3, Fluid Dynamics, Los Alamos National Lab, Los Alamos, NM 87544}
\author[0000-0002-4854-8636]{Roseanne M. Cheng}
\affiliation{T-3, Fluid Dynamics, Los Alamos National Lab, Los Alamos, NM 87544}
\author[0000-0003-1707-7998]{Nicole M. Lloyd-Ronning}
\affiliation{CCS-2, Computational Physics and Methods, Los Alamos National Lab, Los Alamos, NM 87544}
\author{Shane L. Larson}
\affiliation{Department of Physics and Astronomy, CIERA Northwestern University, Evanston, IL 60208}
\author{Gabriel O. Casabona}
\affiliation{Department of Physics and Astronomy, CIERA Northwestern University, Evanston, IL 60208}




\begin{abstract}

We compare the evolution of binary systems evolved in the MESA stellar evolution code to those in the COSMIC population synthesis code.  Our aim is to convey the robustness of the equations that model binary evolution in the COSMIC code, particularly for the cases of high mass stars with closely orbiting compact object companions.  Our larger goal is to accurately model the rates of these systems, as they are promising candidates for the progenitor systems behind energetic, longer lasting, radio bright GRB jets.  These systems also may be key contributors to the rates of binary black hole mergers throughout our universe.
\\ \\

\end{abstract}



\section{Introduction}
\label{sec:intro}
The well developed open-source codes MESA and COSMIC can both be used to model individual binary systems. 
MESA is a 1D+ hydrodynamic stellar evolution code that solves the coupled structure and composition equations simultaneously.
\citep{Paxton2011, Paxton2013, Paxton2015, Paxton2018, Paxton2019, paxton_bill_2021_5798242}\footnote{We use release r21.12.1}.  It has the capability to model binary systems under a wide range of scenarios, with sophisticated mass-transfer schemes and common envelope evolution.
COSMIC (Compact Object Synthesis and Monte Carlo Investigation Code) \citep{Brev20} is a rapid population synthesis code.  It relies on the equations of \cite{HPT00} and \cite{HPT02} to evolve stars in binary systems, and includes extensive updated prescriptions to to account for massive stars and binary interactions \citep{Brev20}.  \\


\begin{figure*} 
    \centering
    \includegraphics[width=\linewidth, height=7.5cm]{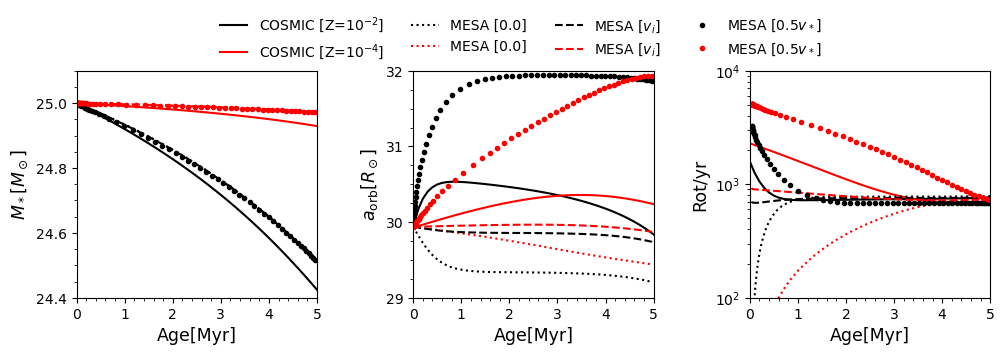}
    \caption{Evolution of mass {\bf left panel}, binary separation {\bf middle panel}, and stellar angular velocity of {\bf right panel} for a variety of COSMIC and MESA simulations involving a $25 M_{\odot}$ main sequence star with a $15 M_{\odot}$ black hole companion with an initial orbital period of 3 days.  The red lines show simulations for the massive star metallicity of $Z=0.1$, while the black lines show simulations for metallicity of $Z=0.0001$.  The solid lines show COSMIC simulations, while the dotted, dashed, and circled lines show MESA simulations for stellar angular velocity of 0, close to COSMIC's initial angular velocity, and $1/2$ the break-up velocity of the star.}
    \label{fig:fig1}
\end{figure*}

Here we compare the evolution of single binary systems within each of these codes, focusing in particular on a massive star with a closely orbiting black hole companion.  Our broad goal is to better constrain the nature of gamma-ray burst progenitors.  In particular, radio bright long gamma-ray bursts shows significantly longer prompt gamma-ray emission and higher isotropic energy compared to radio-dark GRBs. \cite{LR22} conjectured that a massive star in an interacting binary system may experience significant spin up due to its compact object companion and have a circumbinary medium that is denser and less ``wind-like'', compared to the circumburst medium of a single star. As such, these systems may explain the unique properties of radio bright GRBs.  Therefore, we would like to understand the viability of these systems as GRB progenitors and the expected rates of these systems throughout cosmic time.  We note these interacting binary systems also have important implications for constraining rates of black hole binary mergers.\\


We aim to verify that the numerical methods in COSMIC for modeling the evolution of single binary systems capture the physics to the extent we need for robust population statistics.
To this end,  we compare the evolution of single binaries in MESA with that of COSMIC throughout the lifetime of the (second) massive star.  We focus in particular on the star's mass loss, the binary separation, and angular momentum.  The latter is important to our larger investigation, as the angular momentum helps us determine whether the massive star is capable of producing a highly spinning black hole-disk system that can launch a relativistic GRB jet upon collapse.  \\





\section{Evolution of Single Binary in COSMIC and MESA} \label{jet}

 
\subsection{Simulations Setups} 

 Our system of choice is a massive star ($25 M_{\odot}$) with a black hole companion ($15 M_{\odot}$), orbiting with a period of 3 days (a separation of roughly $30 R_{\odot}$).  We investigate the evolution for two different metallicities -  $Z=10^{-2}$ and $Z=10^{-4}$ - where $Z$ is the fraction of elements heavier than helium.  Additionally, we compare the evolution of our MESA binaries for different initial rotational velocities of the massive star.  We refer the reader to the COSMIC\footnote{\url{https://cosmic-popsynth.github.io/docs/stable/index.html}} and MESA\footnote{\url{https://docs.mesastar.org/en/release-r22.05.1/index.html}} documentation for the many simulation initial conditions and flags, but we have chosen the set-ups to be as close as possible between the two codes.  
 We have verified that our results in MESA converge with increasing spatial and temporal resolution.  We have done similar convergence tests in COSMIC, for increasing temporal resolution. 

\subsection{MESA and COSMIC Outputs}
  Figure 1 shows the mass loss (left panel), orbital separation (middle panel) and angular velocity (right panel), for two COSMIC runs and six MESA runs, for the input parameters described above.  We show three distinct sets of MESA simulations that differ by the initial velocity of the massive star in the system - zero angular velocity, an angular velocity that is close to COSMIC's initial angular velocity, and an angular velocity that is $1/2$ the breakup velocity of the star.  \\


{\bf Mass:}  The left panel shows that the mass loss between COSMIC and MESA are very similar with the differences likely due to the more nuanced EOS prescriptions in MESA.  As expected, we see the higher metallicity runs experiencing more mass loss due to a combination of more line driven winds, and a more extended stellar envelope leading to more Roche lobe overflow. \\

{\bf Binary Separation:} 
The middle panel shows the evolution of the binary separation.  We see the results are relatively similar - with binary separation decreasing - between the COSMIC and MESA runs for the case of zero stellar angular velocity and angular velocity that matches COSMIC's initial conditions.  
In contrast, the MESA simulations in which the star has an angular velocity of half of the break-up velocity show an increasing binary separation.  
The evolution of binary separation among all simulations is illuminated by the change in the stellar angular velocity and conserving angular momentum in the system.\\


{\bf Angular Velocity:}  The right panel shows the angular velocity of the star.  We see that the simulation results converge over the $\sim 5$ Myr lifetime of the massive star, to about 
to about the same angular velocity of roughly $10^{3}$ rotations/yr. 







\section{Conclusions} \label{Conclusion}
 We confirm the robustness and reliability of COSMIC's binary evolution prescription, which provides the confidence in using this code as a tool to model populations of binary systems throughout the lifetime of our universe.  
 


\section{Acknowledgements}
\label{sec:ack}
 
We thank Carl Fields for helpful MESA and stellar evolution insights. This work was supported by the US Department of Energy through Los Alamos National Laboratory.  Los Alamos National Laboratory is operated by Triad National Security, LLC, for the National Nuclear Security Administration of U.S. Department of Energy (Contract No. 89233218CNA000001).   Research presented was supported by the Laboratory Directed Research and Development program of LANL project number 20230115ER. We acknowledge LANL Institutional Computing HPC Resources under project w23extremex. 
LA-UR-23-24423

\bibliography{main}{}
\bibliographystyle{aasjournal}



\setlength{\arrayrulewidth}{0.5mm}
\setlength{\tabcolsep}{3pt}
\renewcommand{\arraystretch}{1.3}

\end{document}